\documentclass[aps,print,superscriptaddress,pre,twocolumn]{revtex4}

\usepackage{amsmath,amssymb}
\usepackage{graphicx}
\usepackage{mathrsfs}
\usepackage[dvipsnames]{xcolor}
\usepackage{subfigure}
\usepackage{ulem}
\usepackage{cancel}

\usepackage[utf8]{inputenc}
\usepackage[english]{babel}

\usepackage[
    breaklinks=true,
    colorlinks=true,
    citecolor=blue,
    linkcolor=blue,
    urlcolor=blue
]{hyperref}

\begin{document}

\title{Near-Threshold Kink Trapping Enhances Defect Healing in an Adaptive $\boldsymbol{\phi^4}$ Chain}

\author{Sh. Minapoor}
\affiliation{Department of Textile Engineering, University of Neyshabur, Neyshabur, Iran}

\author{Sergey V. Dmitriev}
\affiliation{National Research University Higher School of Economics, St. Petersburg 190121, Russia}

\author{A. Moradi Marjaneh}
\email[Corresponding author:]{moradimarjaneh@iau.ac.ir, moradimarjaneh@gmail.com}
\affiliation{Department of Physics, Qu.C., Islamic Azad University, Quchan, Iran}
\affiliation{A.I. Alikhanyan National Science Laboratory (Yerevan Physics Institute), Alikhanian Brothers Street 2, 0036 Yerevan, Armenia}

\begin{abstract}
We study the scattering of topological kinks in a discrete $\phi^4$ chain with a localized, adaptive defect, motivated by autonomous self-healing mechanisms in fibrous and polymeric chains. The defect represents damaged flanking bonds whose coupling stiffness evolves dynamically via an internal healing variable, driven by the local energy density and moderated by a viscoelastic relaxation rate $\beta$. Using high-resolution numerical simulations, we identify a sharp transition at a critical impact velocity $v_c$ separating kink reflection from transmission ($v_c \simeq 0.4183$ for the benchmark initial defect stiffness $C_{\min} = 0.2$). While subcritical kinks ($v_i < v_c$) bounce back with little energy deposition and supercritical ones ($v_i > v_c$) cross the defect too quickly to induce noticeable recovery, kinks near the threshold ($v_i \simeq v_c$) undergo resonant trapping. This prolonged dwell time transfers substantial vibrational energy into the defect, driving the healing variable to nearly full restoration ($h_{\max} \approx 0.98$). Notably, while $v_c$ and the peak defect energy $E_{d,\max}$ depend strongly on $C_{\min}$ and are largely unaffected by $\beta$, non-zero relaxation ($\beta > 0$) shortens the trapping window and reduces the final recovery efficiency. Our results show that energy-driven self-healing in discrete lattices is controlled by the cumulative dwell time and time-integrated energy excitation rather than the instantaneous impact peak alone.
\end{abstract}


\maketitle

\section{Introduction}

Self-healing materials have drawn increasing interest in smart fabrics, fiber-reinforced composites and soft wearable devices with potential applications in aerospace and advanced industrial technologies \cite{Chaudhary.2024.Textiles,Kontiza.2024.Polymers,Kausar.2023.JCS,Huynh.2017.AdvMater}. Self-healing in polymeric and fibrous materials can be achieved through intrinsic mechanisms involving reversible covalent bonds \cite{Yu.2018.JMPS,Cao.2024.JAPS,Ratwani.2023.PMS} and noncovalent interactions \cite{Stukalin.2013.Macromolecules,Irzhak.2022.Polymers}, or through extrinsic strategies based on the release of encapsulated healing agents \cite{Irzhak.2022.Polymers,Ramesh.2022.MatToday}. In self-healing thermoplastic polymer fibers, damage repair may proceed through viscoelastic shape recovery that facilitates fracture closure, followed by chain diffusion and interfacial rebonding \cite{Hornat.2020.NatCommun}. Biomimetic fiber systems have also demonstrated the role of interphase morphology and shape-memory behavior in self-healing \cite{Yang.2018.Chem}. However, practical implementation remains challenging, as the molecular mobility required for healing can compromise mechanical strength and stability \cite{Wen.2021.RFP,Wang.2025.Molecules}. Furthermore, healing kinetics and the need for recovery under service conditions impose additional limitations on performance \cite{Wang.2025.Molecules}.

To model self-healing behavior, previous studies have developed continuum damage--healing formulations \cite{Smojver.2022.IJDM,Pan.2018.IJDM,Subramanian.2023.IJSS,Subramanian.2022.MechMater}, constitutive models describing the mechanical behavior of self-healing polymers and composites \cite{Dai.2023.IJMS,Yu.2020.JMPS}, and multiscale approaches for self-healing textile composites \cite{Udhayaraman.2022.MAMS,Zhang.2024.Polymers,Smojver.2022.CompStruct}. Other studies have introduced multiphysics frameworks coupling mechanical degradation with diffusion and chemical reactions \cite{Zhang.2022.IJSS,SanzHerrera.2019.IJSS,Bellezza.2022.JECS,Zhou.2024.CompStruct}, as well as models addressing fiber--matrix interfacial healing \cite{Liu.2023.CBM} and numerical approaches to self-healing cementitious materials \cite{Jefferson.2018.AMI}. These formulations provide valuable insights into damage evolution, healing kinetics, mechanical recovery, and coupled transport processes at the material and structural scales. Consequently, there remains an acute lack of discrete physical models capable of capturing how dynamic energy transport along a single fibrous chain can autonomously trigger and govern local bond restoration.

In condensed matter and other branches of physics, discrete Klein--Gordon and $\phi^4$ chains provide a foundational framework for modeling localized excitations, phase transitions, and energy transport in macromolecular lattices~\cite{Makhankov.PhysRep.1978, Lohe.PRD.1979, Campbell.PhysD.1983, Kevrekidis.2019, CuevasMaraver.2014}. A key feature of these systems is the propagation of topological solitons (kinks) acting as robust, particle-like energy carriers. The scattering of kinks off local inhomogeneities and structural defects exhibits rich behavior, including transmission, reflection, radiation, and resonant trapping mediated by energy exchange with defect modes. In a damaged polymeric lattice, bond rupture naturally corresponds to a localized defect that modifies the inter-site coupling stiffness.

Unlike classical impurity studies that assume static or prescribed defects~\cite{Fei.1992, Fakhretdinov2025419, Ekomasov20261135, Fakhretdinov.2023}, self-healing systems are adaptive: structural restoration is driven by localized mechanical strain and energy input. Motivated by this mechanism, we investigate a minimal discrete $\phi^4$ lattice with an adaptive localized defect as a model for energy-driven self-healing in damaged filaments. The defect coupling evolves dynamically with the local energy density, establishing a feedback loop between solitary wave excitations and structural recovery.

The primary objective of this work is to examine the scattering and trapping dynamics of topological kinks interacting with this adaptive defect. Using high-resolution numerical simulations, we identify a sharp dynamical transition at a critical velocity $v_c$ separating subcritical reflection $(v_i < v_c)$ from supercritical transmission $(v_i > v_c)$. Near this threshold $(v_i \simeq v_c)$, kink trapping leads to prolonged oscillations and resonant defect-mode excitation, maximizing the healing variable toward full restoration. Furthermore, we analyze the parametric dependence of the dynamics on initial damage severity $C_{\min}$ and structural viscoelastic relaxation $\beta$. We show that while $v_c$ is predominantly determined by $C_{\min}$ and remains largely insensitive to $\beta$, viscous dissipation significantly reduces the interaction residence time and suppresses the ultimate healing efficiency $h_{\max}$. This study links discrete nonlinear physics with materials science, offering fundamental insights into soliton-mediated self-healing in adaptive networks.

\section{The Adaptive \texorpdfstring{$\phi^4$}{phi4} Lattice Model}
\label{sec:model}

We model a damaged fibrous filament as a one-dimensional discrete $\phi^4$ chain governed by on-site double-well bistability and inter-site elastic interactions. Let $\phi_n(t)$ denote the scalar displacement at lattice site $n \in \mathbb{Z}$ and time $t$, with uniform lattice spacing $\Delta x$. The equations of motion read
\begin{align}
\ddot{\phi}_n ={}& \frac{C_{n,n+1}(t)\,(\phi_{n+1} - \phi_n) - C_{n-1,n}(t)\,(\phi_n - \phi_{n-1})}{\Delta x^2} \nonumber \\
{}&+ \phi_n - \phi_n^3 - \gamma_n \dot{\phi}_n,
\label{eq:motion}
\end{align}
where $C_{n,n+1}(t)$ represents the effective inter-site coupling strength between sites $n$ and $n+1$ (assuming symmetric coupling parameters for the neighboring bonds such that $C_{n,n+1} = C_{n-1,n} = C_n$ in the homogeneous regions), and the local potential $V(\phi) = \frac{1}{4}(\phi^2 - 1)^2$ exhibits degenerate ground states at $\phi = \pm 1$. The boundary dissipation profile $\gamma_n$ acts as an absorbing layer near the domain edges and is identically zero in the bulk of the lattice. Away from the damaged flanking bonds, the elastic coupling is uniform and static, $C_{n,n+1} = 1$.

\subsection{Adaptive Defect and Coupling Modulation}

To capture localized mechanical damage, such as micro-crack opening or polymeric cross-link rupture, we introduce a localized defect centered at site $n_d$, where the two adjacent bonds flanking the core site have a modified stiffness:
\begin{equation}
C_{n,n+1}(t) = \begin{cases}
C_d(t), & n \in \{n_d - 1, \, n_d\}, \\
1, & \text{otherwise}.
\end{cases}
\label{eq:C_coupling}
\end{equation}

The instantaneous defect coupling $C_d(t)$ evolves dynamically between a damaged baseline and complete structural restoration according to
\begin{equation}
C_d(t) = C_{\min} + \big(1 - C_{\min}\big)\,h(t),
\label{eq:Cd}
\end{equation}
where $C_{\min} \in (0, 1)$ denotes the residual coupling strength of the fractured interface, and $h(t) \in [0, 1]$ is a non-dimensional internal healing variable. The bounds $h = 0$ ($C_d = C_{\min}$) and $h = 1$ ($C_d = 1$) correspond to the completely ruptured and fully restored states, respectively.

\subsection{Energy-Driven Healing Kinetics}

Autonomous self-healing is activated dynamically by local strain and kinetic energy deposition. We formulate the temporal evolution of the internal healing state as an energy-activated nonlinear dynamical process:
\begin{equation}
\frac{\mathrm{d}h}{\mathrm{d}t} = \alpha\,E_d(t)\,(1 - h) - \beta\,h,
\label{eq:dhdt}
\end{equation}

where $\alpha > 0$ governs the rate of energy-induced bond reformation, the saturation factor $(1 - h)$ enforces asymptotic recovery toward the pristine state ($h \to 1$), and $\beta \ge 0$ parameterizes the relaxation rate of the internal state (modeling reversible debonding or structural aging in the absence of sustained excitation). In the dissipationless limit $\beta = 0$, Eq.~\eqref{eq:dhdt} can be integrated directly for an initial state $h(0) = h_0$, yielding
\begin{equation}
h(t) = 1 - (1 - h_0) \exp\left(-\alpha \int_0^t E_d(t')\,\mathrm{d}t'\right),
\label{eq:h_analytic}
\end{equation}
or equivalently,
\begin{equation}
\ln\left[\frac{1 - h(t)}{1 - h_0}\right] = -\alpha \int_0^t E_d(t')\,\mathrm{d}t'.
\label{eq:ln_h_analytic}
\end{equation}
For an initially unhealed defect with $h(0) = 0$, this simplifies to $\ln[1 - h(t)] = -\alpha \int_0^t E_d(t')\,\mathrm{d}t'$. This closed-form relation directly underscores that the asymptotic recovery is determined by the cumulative energy delivered to the defect over time, rather than by instantaneous peak values.

The local defect energy $E_d(t)$ represents the concentrated energy stored at the defect core, defined consistently with the discrete implementation as

\begin{align}
E_d(t) ={}& \frac{1}{2}\dot{\phi}_{n_d}^2 + \frac{1}{4}\big(\phi_{n_d}^2 - 1\big)^2 \nonumber \\
&{}+ \frac{C_d(t)}{2\,\Delta x} \Big[(\phi_{n_d+1} - \phi_{n_d})^2 + (\phi_{n_d} - \phi_{n_d-1})^2\Big].
\label{eq:Ed}
\end{align}

Equation~\eqref{eq:Ed} accounts for the kinetic energy of the core site, the double-well barrier crossing energy, and the strain energy of the adjacent bonds flanking the defect site.

\section{Numerical Results and Discussion}
\label{sec:results}

\subsection{Kink Scattering Dynamics and Energy-Driven Healing}
\label{subsec:scattering_healing}

To elucidate the autonomous response of the damaged fibrous medium under mechanical excitation, we investigate the collision between an incoming topological kink and the adaptive defect located at the lattice center $X=0$. 

The lattice equations of motion are integrated using the velocity-Verlet algorithm on a uniform grid of $N = 400$ sites ($n \in [-200, 200]$) with spatial spacing $\Delta x = 0.1$ and integration time step $\Delta t = 0.005$. To prevent spurious reflections of radiated phonons from entering the scattering region, localized absorbing sponge layers with width $n_{\mathrm{damp}} = 40$ sites and damping coefficient $\gamma_{\max} = 0.05$ are imposed at both boundaries. A single topological kink is initialized at $X_{\mathrm{init}} = -10.0$ and propelled toward the defect core. The interaction duration $t_{\mathrm{interaction}}$ is rigorously recorded as the cumulative dwell time during which the kink center resides within the localized interaction window $|X(t)| \le X_0 = 2.0$.

\begin{figure*}[!ht]
\centering
  \subfigure[]{\includegraphics[width=0.32\textwidth]{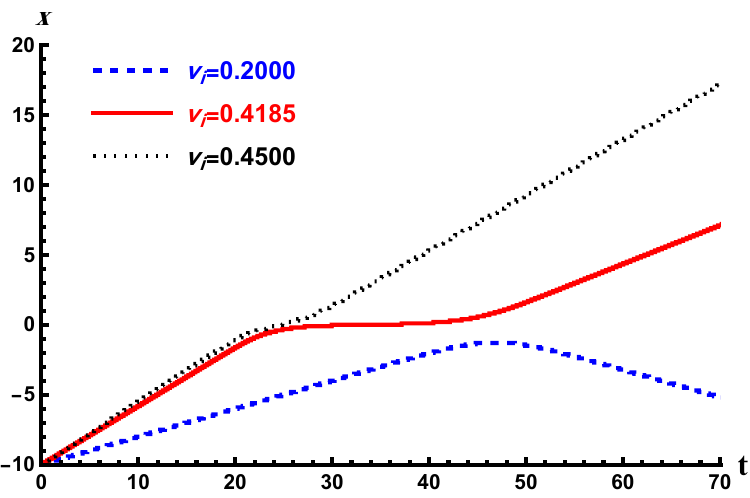}\label{fig:trajectories_x}}
  \subfigure[]{\includegraphics[width=0.32\textwidth]{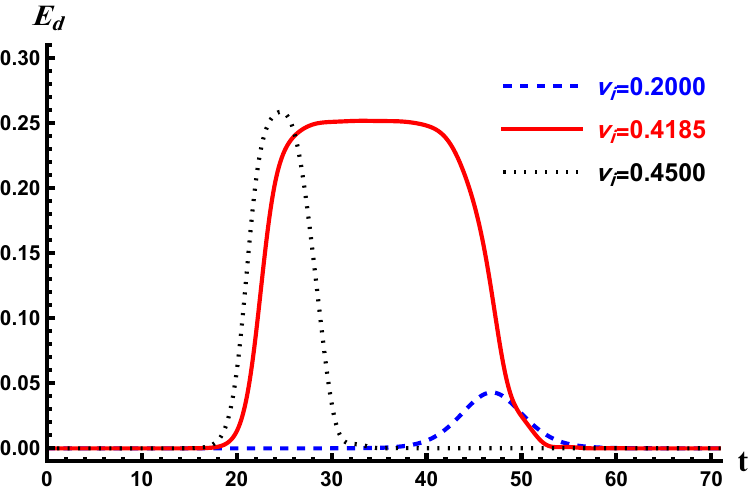}\label{fig:trajectories_Ed}}
  \subfigure[]{\includegraphics[width=0.32\textwidth]{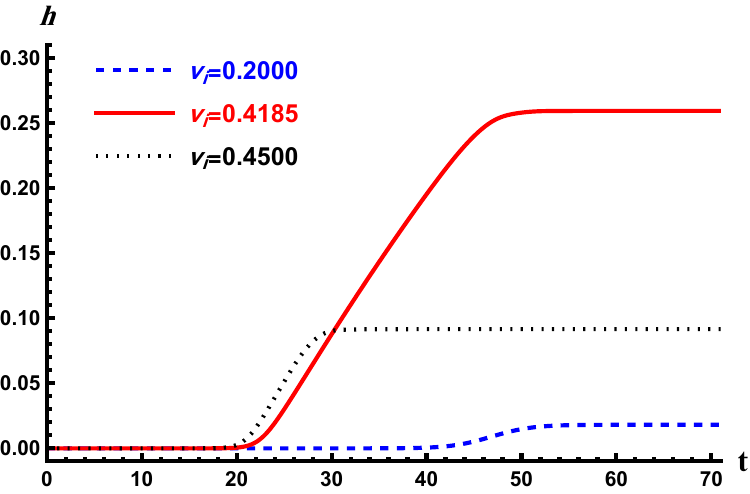}\label{fig:trajectories_h}}
\caption{Synchronized temporal dynamics of the defect interaction for three representative initial kink velocities: subcritical reflection ($v_i = 0.2000$), near-critical resonant trapping ($v_i = 0.4185$), and supercritical transmission ($v_i = 0.4500$), with model parameters fixed at $C_{\min} = 0.2$, $\alpha = 0.05$, and $\beta = 0$. (a) Trajectory of the kink center $X(t)$, (b) localized defect energy density $E_d(t)$, and (c) autonomous self-healing variable $h(t)$.}
\label{fig:triplet_dynamics}
\end{figure*}

Figure~\ref{fig:triplet_dynamics} displays the synchronized time evolution of the kink core trajectory $X(t)$, the localized defect energy $E_d(t)$, and the resulting healing variable $h(t)$ for three representative incident velocities: a subcritical velocity ($v_i=0.2000$), a near-critical velocity ($v_i=0.4185$), and a supercritical velocity ($v_i=0.4500$), with parameters fixed at $C_{\min}=0.2$, $\alpha=0.05$, and $\beta=0$.

A correlated inspection of Figs.~\ref{fig:trajectories_x}--\ref{fig:trajectories_h} reveals three distinct interaction regimes:
\begin{enumerate}
\item \textbf{Reflection regime ($v_i=0.2000$):}
As shown in Fig.~\ref{fig:trajectories_x}, the low-velocity kink decelerates upon approaching the defect core and turns back at approximately $t\simeq 47$--$48$, never crossing $X=0$. The associated defect excitation remains weak (Fig.~\ref{fig:trajectories_Ed}), reaching only a modest peak of about $E_d \simeq 0.04$ near $t\simeq 47$. Consequently, the kinetic driving term in Eq.~\eqref{eq:dhdt} is marginal, and the healing variable increases only slightly, saturating at a near-zero plateau $h \simeq 0.02$ (Fig.~\ref{fig:trajectories_h}).

\item \textbf{Near-critical resonant interaction and delayed transmission ($v_i=0.4185$):}
When the incident velocity approaches the critical boundary, the kink exhibits an extended dwell time in the defect vicinity. In Fig.~\ref{fig:trajectories_x}, the core remains trapped near $X\simeq 0$ over a wide interval $t\approx 25$--$45$ before escaping. This prolonged localization enables sustained energy pumping into the defect degrees of freedom, producing a broad high-energy episode in $E_d(t)$ (Fig.~\ref{fig:trajectories_Ed}) that reaches $E_d \simeq 0.25$. The accumulated excitation activates the healing kinetics most efficiently among the three cases, driving $h(t)$ to its largest plateau, $h \simeq 0.26$, with saturation occurring around $t\approx 45$--$50$ (Fig.~\ref{fig:trajectories_h}). Physically, the strong response originates from resonance-assisted trapping, which maximizes the time-integrated defect excitation entering Eq.~\eqref{eq:dhdt}.

\item \textbf{Direct transmission regime ($v_i=0.4500$):}
For velocities above the critical threshold, the kink rapidly penetrates the defect and crosses $X=0$ at approximately $t\simeq 22$ (Fig.~\ref{fig:trajectories_x}). The defect energy exhibits a sharp, short-lived peak of about $E_d \simeq 0.26$ near $t\simeq 25$ (Fig.~\ref{fig:trajectories_Ed}). However, because the interaction window is brief, the cumulative activation of the healing channel is limited and the recovery saturates at a substantially smaller value, $h \simeq 0.09$, by $t\approx 30$ (Fig.~\ref{fig:trajectories_h}). This highlights a key design principle: large instantaneous energy density alone does not guarantee strong healing; rather, efficient self-repair requires sufficiently long residence near the damaged bonds.
\end{enumerate}

Importantly, since $\beta=0$, the healing dynamics obeys $\dot{h}=\alpha E_d(1-h)\ge 0$, ensuring that the recovery is strictly monotonic. After the kink exits the interaction region and $E_d(t)\to 0$, the healing variable freezes at its attained plateau. The coupled dynamics therefore realizes irreversible autonomous self-healing by harvesting localized energy from nonlinear wave--defect interactions.

\subsection{Resonant Trapping, Critical Velocity, and Interaction Time Anomaly}
\label{sec:resonant_trapping}

\begin{figure*}[!ht]
\centering
  \subfigure[]{\includegraphics[width=0.32\textwidth]{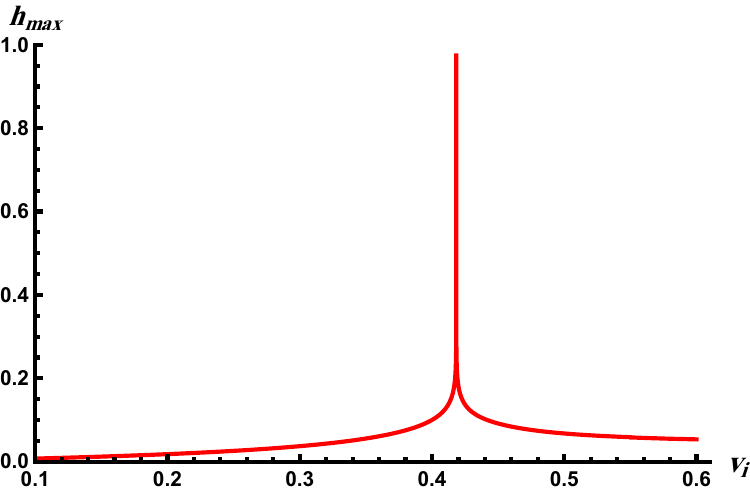}\label{fig:hmax_scan}}
  \subfigure[]{\includegraphics[width=0.32\textwidth]{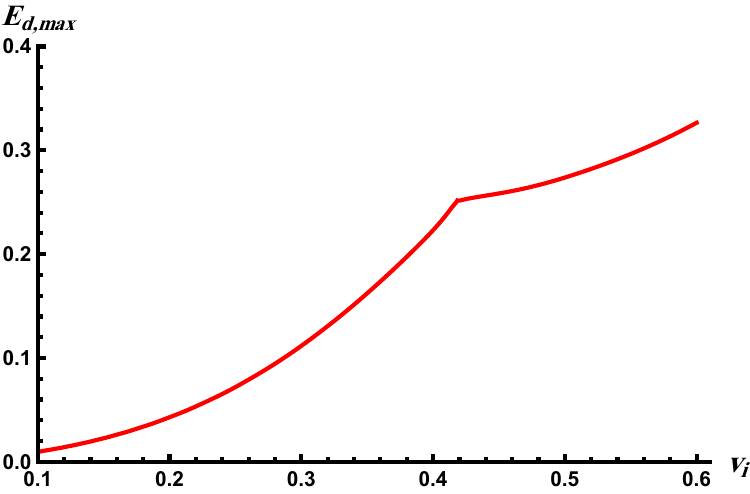}\label{fig:Edmax_scan}}
  \subfigure[]{\includegraphics[width=0.32\textwidth]{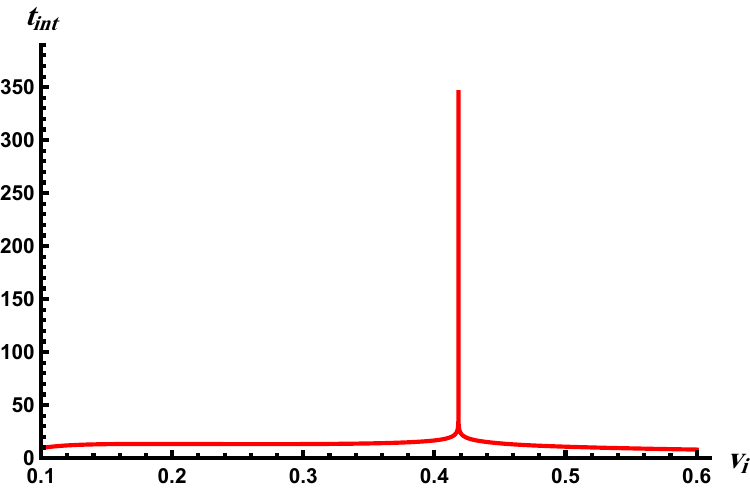}\label{fig:t_interaction_scan}}
\caption{Dynamical response of the defect system as a function of the initial kink velocity $v_i$ for $\alpha = 0.05$, $\beta = 0$, and $C_{\min} = 0.2$: 
(a) maximum self-healing variable $h_{\max}$, 
(b) peak localized defect energy density $E_{d,\max}$, and 
(c) total interaction duration $t_{\mathrm{interaction}}$. 
A pronounced resonant transition emerges at the critical threshold $v_c \simeq 0.41834643$, characterized by a singular anomaly in the dwell time and a near-complete lattice restoration peak, contrasted with a smooth, monotonic rise in peak energy injection.}
\label{fig:hmax_Edmax_tinter}
\end{figure*}

To comprehensively elucidate the parametric dependence of the lattice response on the incident kink velocity, we perform fine-resolution velocity scans. Figure~\ref{fig:hmax_Edmax_tinter} summarizes the asymptotic response of the defect system as a function of the initial velocity $v_i$ for baseline parameters $\alpha = 0.05$, $\beta = 0$, and $C_{\min} = 0.2$.

The response curves across Figs.~\ref{fig:hmax_scan}--\ref{fig:t_interaction_scan} manifest three distinct kinematic regions demarcated by a sharp transition:
\begin{enumerate}
    \item \textbf{Subcritical Reflection Zone ($v_i < v_c$):} For low-to-moderate initial velocities, the kink’s kinetic energy is insufficient to surmount the effective potential created by the weakened coupling and radiative defect mode exchange. Consequently, the kink reflects elastically off the defect.
    Consequently, the kink reflects off the defect. Within this window, $h_{\max}$ remains at a modest baseline ($h_{\max} \lesssim 0.05$ for $v_i \lesssim 0.3$, rising gradually to $\sim 0.15$ near $v_i \approx 0.40$), and the dwell time remains short ($t_{\mathrm{interaction}} \sim 10$--$20$), reflecting an elastic-like impulsive collision. Correspondingly, the localized peak energy density $E_{d,\max}$ increases smoothly from near zero up to $\sim 0.24$.
    
    \item \textbf{Resonant Trapping Anomaly ($v_i \to v_c \simeq 0.41834643$):} Approaching the critical velocity in the non-dissipative limit, the system undergoes an intense dynamic transition. As depicted in Fig.~\ref{fig:t_interaction_scan}, the interaction duration $t_{\mathrm{interaction}}$ undergoes a near-divergent spike, reaching $t_{\mathrm{interaction}} \approx 350$. This divergence indicates that the incoming topological excitation is captured by the defect's internal modes, executing multiple localized oscillations within the damaged zone before escaping. This prolonged localization serves as an effective energy accumulator, channeling sustained kinetic activation into the local healing channel. Consequently, $h_{\max}$ exhibits a sharp resonant peak approaching complete restoration ($h_{\max} \approx 0.98$) as shown in Fig.~\ref{fig:hmax_scan}. Notably, Fig.~\ref{fig:Edmax_scan} shows that $E_{d,\max}(v_i)$ does not form a localized peak, but merely exhibits a minor slope adjustment near $v_c$ ($E_{d,\max} \approx 0.26$), confirming that high healing yields are mediated primarily by the extended interaction dwell time rather than an anomalous spike in instantaneous peak energy.
    
    \item \textbf{Supercritical Transmission ($v_i > v_c$):} For velocities exceeding the critical threshold, the kink overcomes the effective potential barrier and promptly crosses the defect. While higher initial velocities naturally impart greater total kinetic energy—leading to a strictly monotonic rise in peak defect energy up to $E_{d,\max} \approx 0.33$ at $v_i = 0.60$ (Fig~\ref{fig:Edmax_scan})—the dwell time inside the interaction zone drops precipitously back to small values ($t_{\mathrm{interaction}} < 25$, descending toward $\sim 10$ at $v_i = 0.60$). Because autonomous healing kinetics relies fundamentally on cumulative localized excitation, this extremely brief transit window severely curbs the cumulative integral $\int E_d\,dt$, causing $h_{\max}$ to drop sharply to negligible levels ($h_{\max} \lesssim 0.06$ for $v_i \gtrsim 0.50$).
\end{enumerate}
\subsection{Influence of Defect Severity on Resonant Trapping and Healing Efficiency}
\label{subsec:cmin_influence}

To systematically determine how the initial structural degradation of the fibrous substrate affects the activation threshold and ultimate recovery, we examine the dependence of both the resonant velocity $v_c$ and the peak healing response $h_{\max}$ on the minimum defect coupling stiffness $C_{\min}$, as depicted in Fig.~\ref{fig:Different_beta}. The parameter $C_{\min}$ serves as a direct proxy for fiber structural integrity: small values ($C_{\min} \to 0$) represent severe structural tearing or thread detachment, whereas larger values ($C_{\min} \to 1$) correspond to localized minor micro-damage.

As shown in Fig.~\ref{fig:Different_beta}, the critical velocity required to trigger effective self-healing decreases monotonically from $v_c \approx 0.60$ at small $C_{\min}$ to lower values as $C_{\min}$ increases. Physically, a deep defect (low $C_{\min}$) establishes a strong local barrier that reflects moderate-speed excitations, requiring a sufficiently fast and energetic kink to penetrate into the defect core. Once trapped, the kink's localized mode is excited, transferring vibrational energy into the adaptive field $h$ before the kink either escapes or remains permanently pinned. Conversely, in mildly degraded fibers (higher $C_{\min}$), the potential barrier is lower, allowing slow-moving mechanical solitons to enter the defect site, excite localized vibrations, and induce healing.

Simultaneously, the peak healing index $h_{\max}$ displays an overall increasing trend with $C_{\min}$, climbing from approximately $0.62$ in severely damaged lattices toward nearly full structural restoration ($h_{\max} \to 1.0$) as $C_{\min}$ approaches unity. This dual dependence demonstrates an advantageous operational window: while heavily damaged textiles necessitate stronger mechanical perturbations to penetrate the defect core and yield moderate recovery, milder micro-fissures can be autonomously repaired with high efficiency by low-amplitude background vibrational solitons.
\subsection{Spatiotemporal Signatures of Resonant Energy Localization}
\label{subsec:spatiotemporal_maps}

\begin{figure*}[!ht]
\centering
  \subfigure[]{\includegraphics[width=0.48\textwidth]{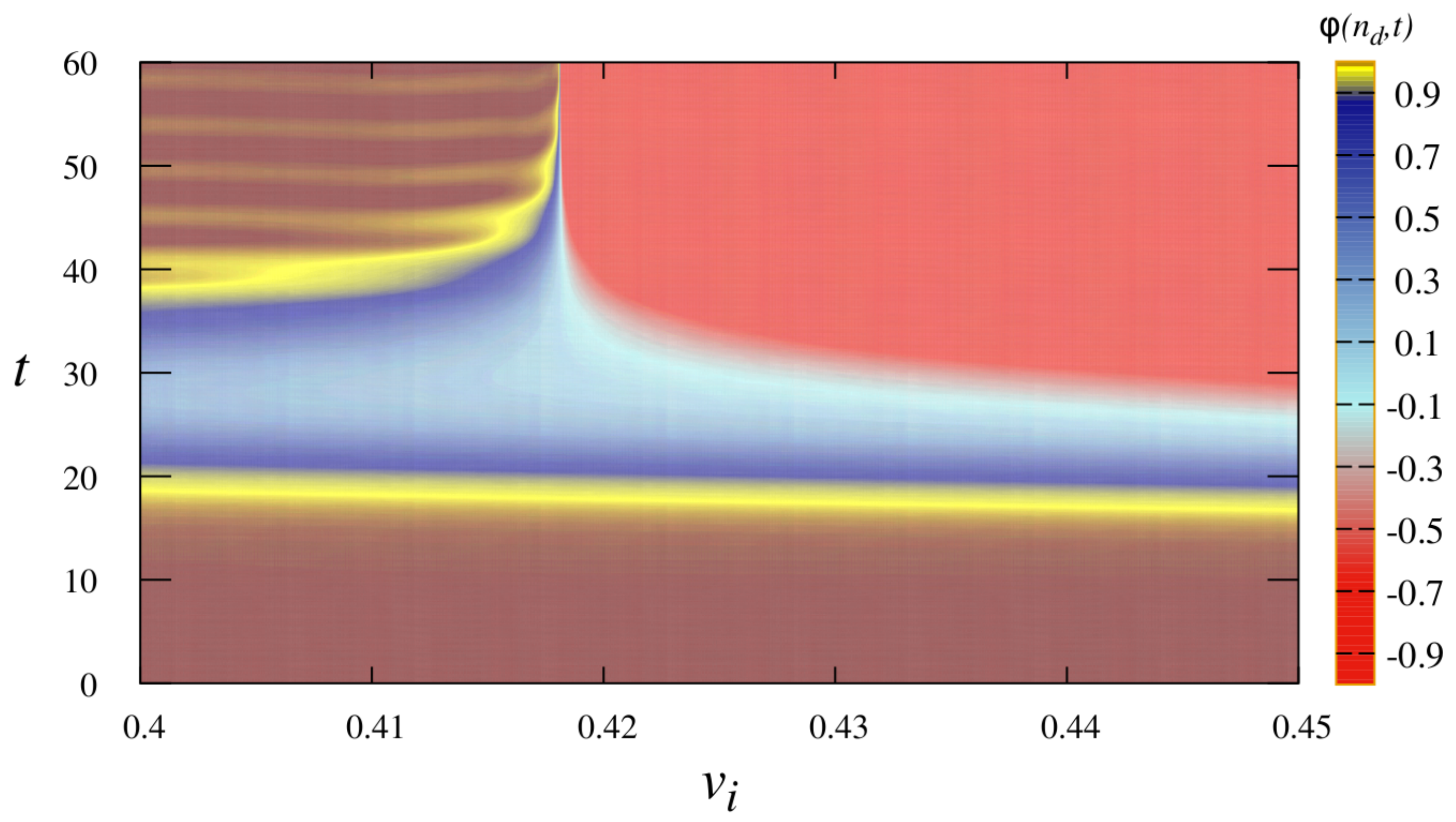}\label{fig:field_xt_v0_alpha_005_beta_0_cmin02}}
  \subfigure[]{\includegraphics[width=0.48\textwidth]{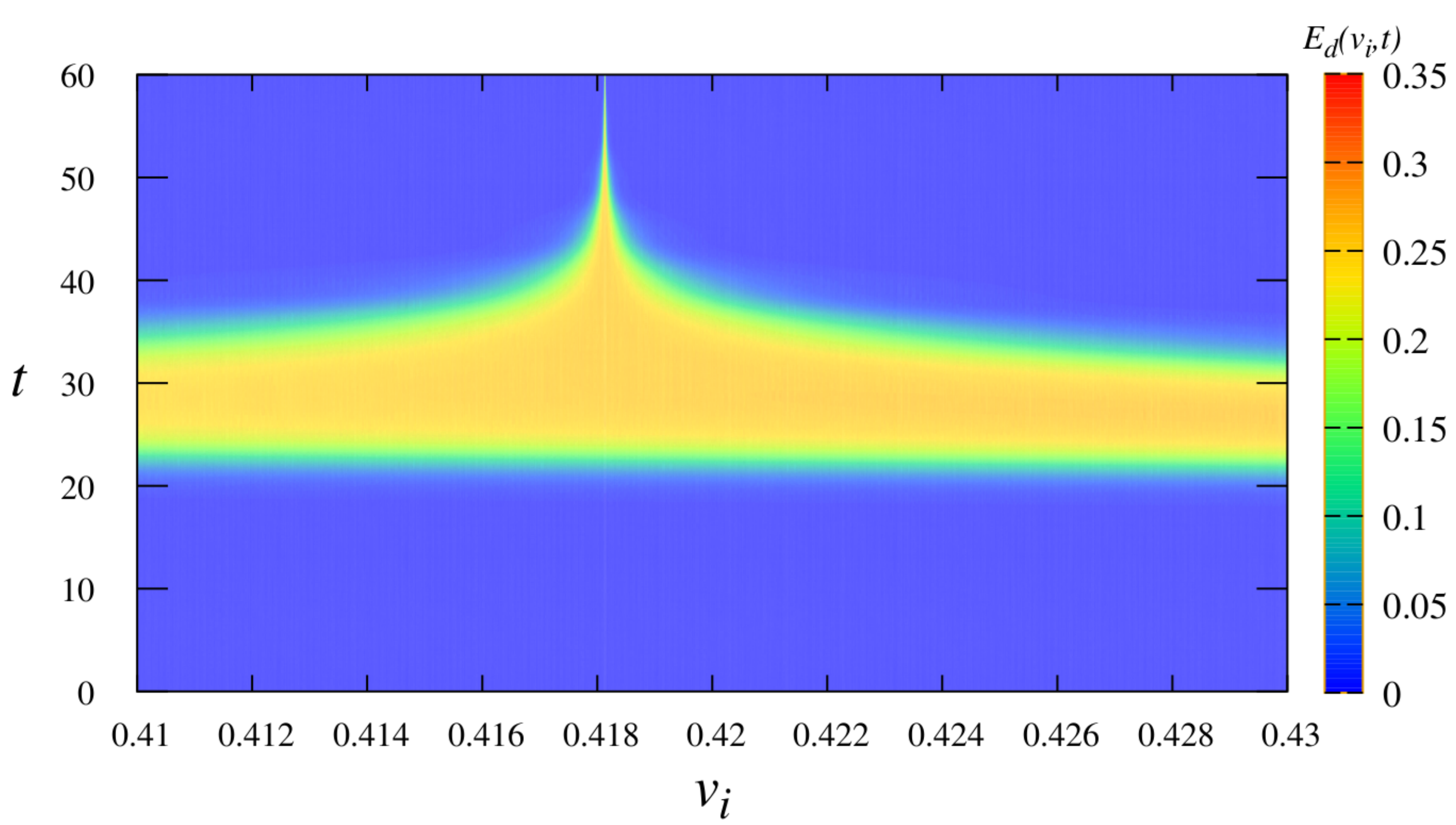}\label{fig:Ed_xt_v0_alpha_005_beta_0_cmin02}}
\caption{Two-dimensional interaction maps in the $(v_i, t)$ parameter space for $\alpha = 0.05$, $\beta = 0$, and $C_{\min} = 0.2$: 
(a) temporal profile of the field at the defect core $\phi(n_d, t)$, and 
(b) localized defect energy density $E_d(n_d, t)$. 
The prominent resonant horn centered precisely at $v_c \simeq 0.41834643$ highlights the prolonged trapping and internal vibrational excitation that fuels autonomous self-healing.}
\label{fig:heatmaps}
\end{figure*}

To uncover the microscopic mechanism behind the dwell-time anomaly and the sharp restoration peak observed near the critical velocity, we construct two-dimensional spatiotemporal heatmaps of the field configuration at the defect site $\phi(n_d, t)$ and the localized defect energy density $E_d(n_d, t)$ across a continuous spectrum of initial velocities $v_i \in [0.2, 0.6]$, as displayed in Fig.~\ref{fig:heatmaps}.

In Fig.~\ref{fig:field_xt_v0_alpha_005_beta_0_cmin02}, the field profile demarcates the sharp kinematic boundary separating kink reflection (where $\phi(n_d, t)$ returns to the vacuum state $\phi = -1$) and kink transmission (where $\phi(n_d, t)$ transitions permanently to $+1$). At the critical boundary $v_i \to v_c$, this interface develops a distinct resonant horn characterized by multiple large-amplitude field oscillations trapped inside the defect well.

Correspondingly, Fig.~\ref{fig:Ed_xt_v0_alpha_005_beta_0_cmin02} maps the localized energy accumulation $E_d(n_d, t)$. Outside the critical neighborhood, the energy density forms a narrow transient pulse of duration $\Delta t \sim 15$--$25$. However, within the resonant corridor around $v_c$, the energy localization undergoes dramatic temporal broadening, persisting over hundreds of time units. This directly corroborates our core thesis: the autonomous healing kinetics governed by Eq.~\eqref{eq:dhdt} is fundamentally an energy-integrating process, where the extended dwell time within the resonant horn acts as a temporal amplifier for bond restoration.

\subsection{Effect of Viscoelastic Relaxation and Structural Degradation}
\label{subsec:beta_influence}

\begin{figure*}[!ht]
\centering
  \subfigure[]{\includegraphics[width=0.32\textwidth]{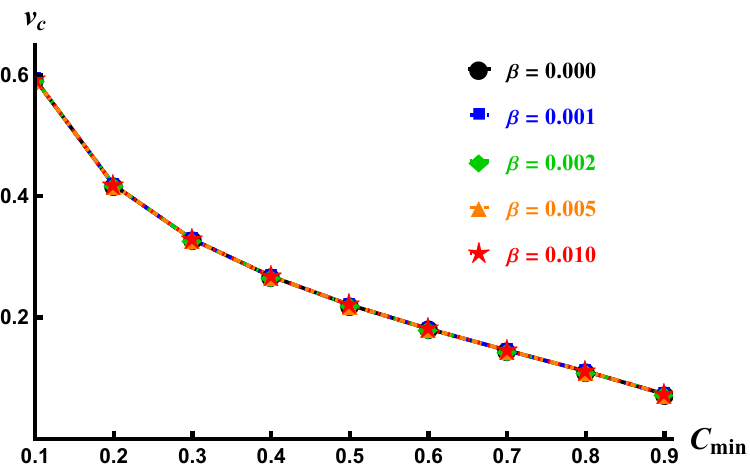}\label{fig:Dif_beta_Vcr}}
  \subfigure[]{\includegraphics[width=0.32\textwidth]{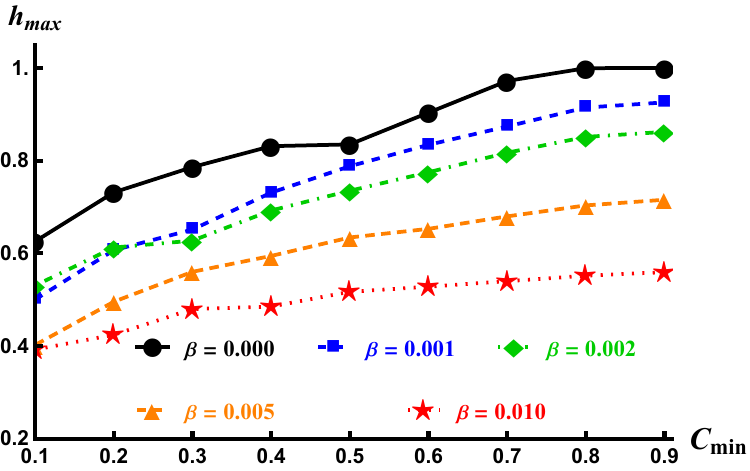}\label{fig:Dif_beta_hmax}}
  \subfigure[]{\includegraphics[width=0.32\textwidth]{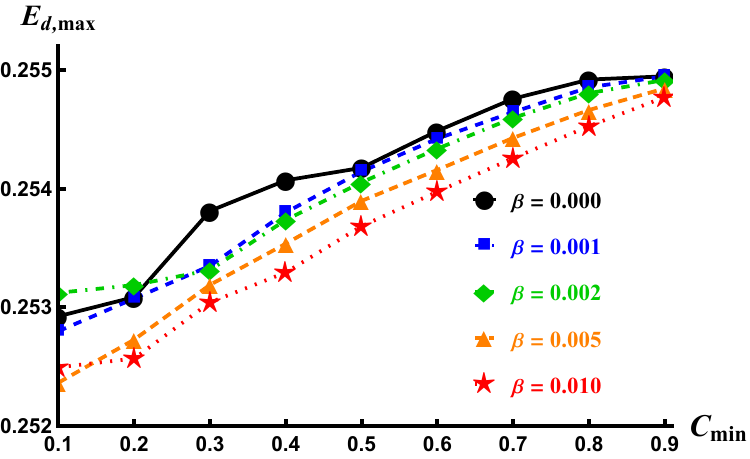}\label{fig:Dif_beta_Edmax}}
\caption{Parametric influence of initial defect severity $C_{\min}$ and viscoelastic relaxation rate $\beta \in [0, 0.010]$ on the resonant scattering dynamics: 
(a) critical impact velocity $v_c$, 
(b) maximum achievable self-healing variable $h_{\max}$, and 
(c) peak localized defect energy density $E_{d,\max}$, with energy pumping rate fixed at $\alpha = 0.05$. While the peak instantaneous energy injection remains virtually constant with less than $1\%$ variation ($E_{d,\max} \in [0.253, 0.255]$), increasing structural relaxation $\beta$ suppresses healing growth, transitioning $h_{\max}$ from a strong $C_{\min}$-dependent ascent at $\beta=0$ to a flattened, $C_{\min}$-independent lower bound at higher $\beta$.}
\label{fig:Different_beta}
\end{figure*}

To evaluate the robustness of the autonomous self-healing mechanism under realistic material degradation conditions, we extend our analysis by incorporating non-zero viscoelastic relaxation rates, $\beta \in [0, 0.010]$, in Eq.~\eqref{eq:dhdt}. Figure~\ref{fig:Different_beta} illustrates the combined parametric influence of the initial defect severity $C_{\min}$ and the structural relaxation rate $\beta$ on the critical impact velocity $v_c$, the maximum achievable healing variable $h_{\max}$, and the peak defect energy density $E_{d,\max}$.

As demonstrated in Fig.~\ref{fig:Dif_beta_Vcr}, the critical velocity $v_c$ exhibits a strong dependence on $C_{\min}$, decreasing monotonically from higher values toward near-zero as $C_{\min}$ approaches $1$. The introduction of a positive relaxation rate ($\beta > 0$) introduces a competitive loss channel against energy-driven bond reformation, slightly modulating the exact threshold required for resonant capture.

More profoundly, Figs.~\ref{fig:Dif_beta_hmax} and \ref{fig:Dif_beta_Edmax} reveal the distinct energetic trade-offs governed by $\beta$. In Fig.~\ref{fig:Dif_beta_Edmax}, the peak instantaneous energy injection $E_{d,\max}$ remains remarkably robust against variations in the relaxation rate, strictly confined within the narrow range of $0.253\text{--}0.255$ (representing a variation of less than $1\%$ across all tested $\beta$ and $C_{\min}$). This remarkable invariance demonstrates that $E_{d,\max}$ is solely dictated by the violent initial localized collision of the kink, well before cumulative viscoelastic dissipation can manifest.

In sharp contrast, Fig.~\ref{fig:Dif_beta_hmax} elucidates the critical interplay between relaxation losses and defect severity on the asymptotic restoration efficiency $h_{\max}$. In the dissipationless regime ($\beta = 0$), $h_{\max}$ exhibits a strong, monotonic dependence on the initial damage parameter, rising markedly from approximately $0.6$ at severe defects ($C_{\min} = 0.1$) to near-complete recovery ($h_{\max} \approx 1$) as $C_{\min} \to 1$. However, as the viscoelastic relaxation rate increases ($\beta > 0$), this dynamic growth is severely penalized. For elevated $\beta$, continuous dissipation ($\beta h$) drains the localized field energy faster than it can accumulate, causing $h_{\max}$ to flatten out into a low, practically horizontal profile that becomes essentially independent of $C_{\min}$. Consequently, highly efficient autonomous repair requires the energy pumping rate ($\alpha E_d$) to decisively outpace intrinsic relaxation losses ($\beta h$), establishing a strict operational threshold for smart textile design against permanent mechanical fatigue.

\section{Conclusion}
\label{sec:conclusion}

In this study, we formulated and investigated a discrete nonlinear $\phi^4$ lattice framework featuring an adaptive, energy-activated localized defect to model the autonomous self-healing and damage-mitigation dynamics of smart fibrous and polymeric media. By coupling the continuous mechanical propagation of topological kinks directly with a localized stiffness-recovery feedback mechanism, we systematically evaluated how impact velocity, interaction duration, initial defect severity ($C_{\min}$), and viscoelastic relaxation ($\beta$) dictate the overall repair and energy-scattering landscape.

Our analysis reveals that kink--defect interactions are strictly partitioned by a well-defined critical impact velocity, determined to be $v_c \simeq 0.41834643$ for the baseline parameters ($C_{\min}=0.2, \alpha=0.05, \beta=0$). While subcritical solitons ($v_i < v_c$) undergo repulsive elastic reflection and supercritical ones ($v_i > v_c$) rapidly traverse the defect with negligible coupling, an anomalous resonant trapping window forms in the immediate vicinity of $v_c$, leading to prolonged energy localization and persistent shape-mode oscillations.

Crucially, our comparative parametric diagnostics uncover a fundamental physical distinction: whereas the peak instantaneous energy injected into the defect ($E_{d,\max}$) increases smoothly and monotonically across the velocity spectrum without local resonance, the interaction dwell time ($t_{\mathrm{interaction}}$) exhibits a dramatic surge, increasing from roughly $15$ to over $350$. This extensive temporal trapping provides the necessary cumulative energy pumping to drive the adaptive healing variable toward a near-complete restoration plateau ($h_{\max} \approx 0.98$). Consequently, autonomous self-healing is fundamentally governed by the time integral of localized excitation rather than the instantaneous impact amplitude.

These dynamics are further illuminated by two-dimensional $(v_i, t)$ spatiotemporal heatmaps of the defect displacement $\phi(n_d, t)$ and energy density $E_d(n_d, t)$, which vividly illustrate the emergence of characteristic resonant ``horns'' anchored at $v_c$. These structures demarcate the temporal boundaries where incoming kinetic energy is efficiently converted into internal defect excitation and subsequent structural recovery. 

Furthermore, a comprehensive scan over initial defect severity ($C_{\min}$) and viscoelastic relaxation rates ($\beta \in [0, 0.010]$) demonstrates that the required resonant activation velocity decreases monotonically as structural integrity increases, dropping from $v_i \approx 0.60$ at $C_{\min} = 0.1$ toward near-zero as $C_{\min} \to 1$. While peak instantaneous energy injection remains virtually invariant with respect to $\beta$, internal structural relaxation introduces a competitive loss channel that noticeably suppresses the asymptotic restoration efficiency ($h_{\max}$). Thus, highly efficient autonomous repair requires energy pumping ($\alpha E_d$) to decisively outpace intrinsic relaxation losses ($\beta h$).

Ultimately, these findings establish a robust theoretical foundation for harnessing nonlinear solitary waves to actuate autonomous structural repair and fatigue mitigation in next-generation smart textiles and macromolecular materials. Future extensions of this work will explore multi-kink collisions, higher-dimensional woven lattice architectures, and the finite-temperature stochastic stability of the self-healing process.

\section*{Acknowledgments}

For SVD this work was supported by the project ‘‘FR-2025-75’’, carried out within the framework of the Basic Research Program at HSE University.

\end{document}